\documentclass[aps,prl,twocolumn,nofootinbib]{revtex4-1}

\usepackage{graphicx}

\begin{document}

\title{Eliminating the Renormalization Scale Ambiguity for Top-Pair Production Using the Principle of Maximum Conformality}

\author{Stanley J. Brodsky$^{1}$}
\email[email:]{sjbth@slac.stanford.edu}

\author{Xing-Gang Wu$^{1,2}$}
\email[email:]{wuxg@cqu.edu.cn}

\address{$^{1}$ SLAC National Accelerator Laboratory, 2575 Sand Hill Road, Menlo Park, CA 94025, USA\\
$^{2}$ Department of Physics, Chongqing University, Chongqing 401331, P.R. China}

\date{\today}

\begin{abstract}

It is conventional to choose a typical momentum transfer of the process as the renormalization scale and take an arbitrary range to estimate the uncertainty in the QCD prediction. However, predictions using this procedure depend on the renormalization scheme, leave a non-convergent renormalon perturbative series, and moreover, one obtains incorrect results when applied to QED processes. In contrast, if one fixes the renormalization scale using the Principle of Maximum Conformality (PMC), all non-conformal $\{\beta_i\}$-terms in the perturbative expansion series are summed into the running coupling, and one obtains a unique, scale-fixed, scheme-independent prediction at any finite order. The PMC scale $\mu^{\rm PMC}_R$ and the resulting finite-order PMC prediction are both to high accuracy independent of the choice of initial renormalization scale $\mu^{\rm init}_R$, consistent with renormalization group invariance. As an application, we apply the PMC procedure to obtain NNLO predictions for the $t\bar{t}$-pair production at the Tevatron and LHC colliders. The PMC prediction for the total cross-section $\sigma_{t\bar{t}}$ agrees well with the present Tevatron and LHC data. We also verify that the initial scale-independence of the PMC prediction is satisfied to high accuracy at the NNLO level: the total cross-section remains almost unchanged even when taking very disparate initial scales $\mu^{\rm init}_R$ equal to $m_t$, $20\,m_t$, $\sqrt{s}$. Moreover, after PMC scale setting, we obtain $A_{FB}^{t\bar{t}} \simeq 12.5\%$, $A_{FB}^{p\bar{p}} \simeq 8.28\%$ and $A_{FB}^{t\bar{t}}(M_{t\bar{t}}>450 \;{\rm GeV}) \simeq 35.0\%$. These predictions have a $1\,\sigma$-deviation from the present CDF and D0 measurements; the large discrepancy of the top quark forward-backward asymmetry between the Standard Model estimate and the data are thus greatly reduced. \\

\begin{description}

\item[PACS numbers] 12.38.Aw, 14.65.Ha, 11.15.Bt, 11.10.Gh

\end{description}

\end{abstract}

\maketitle

Physical predictions in Quantum Chromodynamics (QCD) are in principle invariant under any choice of renormalization scale and renormalization scheme. It is common practice to simply guess a renormalization scale $\mu_R = Q$, $Q$ being a typical momentum transfer of the process, and then vary it over the range $[Q/2, 2\,Q]$. However, this arbitrary procedure leads to scheme-dependent predictions at any finite order in perturbation theory. In fact, a principal ambiguity in perturbative QCD calculations lies in the choice of $\mu_R$. It has been considered as a main systematic error in QCD perturbative analyses.

The Brodsky-Lepage-Mackenzie method (BLM)~\cite{blm} and the Principle of Maximum Conformality (PMC)~\cite{pmc1,pmc2} provide a solution to this problem. The PMC provides the principle underlying BLM scale setting; the BLM is equivalent to PMC through the PMC - BLM correspondence principle~\cite{pmc2}, so we shall treat them on equal footing. When one applies the PMC, all non-conformal $\{\beta_i\}$-terms in the perturbative expansion are summed into the running coupling so that the remaining terms in the perturbative series are identical to that of a conformal theory, i.e., the corresponding theory with $\{\beta_i\} \equiv \{0\}$.

The PMC coefficients and PMC scales may be different under different renormalization schemes, however their combined result will be the same, since the scheme-dependent PMC scales for different schemes are related by commensurate scale relations~\cite{stanlu}. Thus, QCD predictions using PMC are independent of the renormalization scheme. After PMC scale setting, the divergent ``renormalon" series with $n!$-growth disappear, so that a more convergent perturbative series is obtained.

The PMC method satisfies all self-consistency conditions, including the existence and uniqueness of the scale, reflexivity, symmetry and transitivity~\cite{selfconsistence}. In the Abelian limit $N_C \to 0$ at fixed $\alpha=C_F \alpha_s$ with $C_F=(N_c^2-1)/2N_c$, it agrees with the Gell Mann-Low procedure for setting the scale in QED~\cite{qedlimit,gml}. Thus as in QED, the renormalization scale can be unambiguously set at each finite order by the PMC. The PMC scales and coefficients can be set order-by-order. A systematic, scheme-independent procedure for setting PMC scales up to next-to-next-to-leading-order (NNLO) has been presented in Ref.~\cite{pmc2}.

Formally, one needs to choose an initial renormalization scale $\mu^{\rm init}_R$ for PMC. However, the final result when summing all $\{\beta_i\}$-terms to all orders will be independent of $\mu^{\rm init}_R$; i.e. for any observable ${\cal O}$, ${\partial {\cal O}\left(\mu^{\rm PMC}_R\right)}/{\partial \mu^{\rm init}_{r}}\equiv 0$, where $\mu^{\rm PMC}_R$ stands for the PMC scale. This is the invariance principle used to derive renormalization group results such as the Callan-Symanzik equations~\cite{cs}. The PMC scales in higher orders take the form of a perturbative series in $\alpha_s$ so as to properly absorb all $\{\beta_i\}-$ dependent terms associated with renormalization into the $\alpha_s$-running coupling~\cite{pmc2,stanlu}. At fixed order, there is some residual initial-scale dependence because of the unknown-higher-order $\{\beta_i\}$-terms. However, such residual renormalization scale-uncertainty will be greatly suppressed since those higher order $\{\beta_i\}$-terms will be absorbed into the PMC scales' higher-order $\alpha_s$-terms.

As an important application of the PMC, we shall predict the $t\bar{t}$-pair hadroproduction cross-section $\sigma_{t \bar{t}}$ up to NNLO. It has been measured at the Tevatron and LHC with high precision~\cite{cdf,d0,atlas,cms}. Theoretically, $\sigma_{t \bar{t}}$ has been calculated up to NLO within the $\overline{MS}$-scheme~\cite{nason1}. Large logarithmic corrections associated with the soft gluon emission have been investigated and resummed up to next-to-next-to-leading-logarithmic corrections~\cite{moch3}. Even though complete NNLO fixed-order results are not available, parts of the fixed-order NNLO results have been derived through resummation~\cite{moch1}. These results provide the foundation for estimating the NNLO results.

The hadronic cross-section for top quark pair production can be written as:
\begin{equation}
\sigma_{t\bar{t}} = \sum_{i,j} \int\limits_{4m^2_{t}}^{S}\, ds \,\, {\cal L}_{ij}(s, S, \mu_f) \hat \sigma_{ij}(s,\alpha_s(\mu_R),\mu_R,\mu_f) ,
\end{equation}
where the parton luminosity
\begin{displaymath}
{\cal L}_{ij} = {1\over S} \int\limits_s^S {d\hat{s}\over \hat{s}} f_{i/H_1}\left(x_1,\mu_f\right) f_{j/H_2}\left(x_2,\mu_f\right)
\end{displaymath}
with $x_1= {\hat{s} / S}$ and $x_2= {s / \hat{s}}$. Here $S$ denotes the hadronic center-of-mass energy squared and $s=x_1 x_2 S$ is the subprocess center-of-mass energy squared. The parameters $\mu_R$ and $\mu_f$ denote the renormalization and factorization scales, and the functions $f_{i/H_{1,2}}(x_\alpha,\mu_f)$ ($\alpha=1,2$) are the parton distribution functions (PDFs) describing the probability to find a parton of type $i$ with a momentum fraction between $x_\alpha$ and $x_{\alpha} +dx_{\alpha}$ in the hadron $H_{1,2}$. The top quark mass $m_{t}$ is the mass renormalized in the on-shell scheme.

The partonic subprocess cross-sections $\hat \sigma_{ij}$ can be decomposed in terms of the dimensionless scaling-functions $f^m_{ij}$, where $(ij) = \{(q{\bar q}), (gg), (gq), (g\bar{q})\}$ stands for the four production channels and $m=0,1,2$ stands for LO, NLO and NNLO functions respectively. The analytical expressions for $f_{ij}^{0,1,2}(\rho,Q)$ which contain the explicit factorization and renormalization scale dependence can be directly read from the HATHOR program~\cite{hathor}. Up to NNLO, $\hat \sigma_{ij}$ takes the following form
\begin{equation}
\hat\sigma_{ij} = \frac{1}{m^2_{t}}\sum_{m=0}^{2}f_{ij}^{m}(\rho,Q) a^{2+m}_s(Q) \ ,
\end{equation}
where $\rho=4m_t^2/s$ and $a_s(Q)=\alpha_s(Q)/\pi$. There is uncertainty in setting the factorization scale $\mu_f$ which appears even in conformal theory, and its determination is a completely separate issue from the renormalization scale setting. To keep our attention on the renormalization scale, we implicitly set $\mu_f\equiv m_t$. For the initial value of $\mu_R= \mu^{\rm init}_{R}$, we take $\mu^{\rm init}_{R}=Q$, where $Q$ stands for the typical momentum transfer of the process. For example, $Q$ can be taken as $m_t$, $2m_t$, $\sqrt{s}$, etc. As the default choice, we take $Q=m_t$.

According to the PMC, we need to identify the $n^{(1,2)}_f$-dependent terms associated with renormalization. Coulomb-type corrections are enhanced by factors of $\pi$ and the PMC scales can be relatively soft for heavy quark velocity $v = \sqrt{1-4 m^2_t/s} \to 0$. Thus the terms which are proportional to $(\pi/v)$ or $(\pi/v)^2$ have a separate PMC scale and will thus be treated separately~\cite{brodsky1}. More explicitly, the NLO and NNLO scaling-functions can be written as
\begin{eqnarray}
f_{ij}^{1}(\rho,Q) &=& \left[A_{1ij} + B_{1ij} n_f \right] + D_{1ij} \left(\frac{\pi}{v}\right) \ , \\
f_{ij}^{2}(\rho,Q) &=& \left[A_{2ij} + B_{2ij} n_f + C_{2ij} n^2_f\right] +\nonumber\\
&& \left[D_{2ij} +E_{2ij} n_f \right] \left(\frac{\pi}{v}\right) + F_{2ij}\left(\frac{\pi}{v}\right)^2 .
\end{eqnarray}

The PMC scales can be set order-by-order and the final result is
\begin{eqnarray}
m_t^2 \hat\sigma_{ij} &=& A_{0ij} a^2_s(Q_1^*) + \left[\tilde{A}_{1ij}\right] a^3_s(Q_1^{**}) + \left[\tilde{\tilde{A}}_{2ij} \right] a^4_s(Q_1^{**})\nonumber\\
&& + \left(\frac{\pi}{v}\right) D_{1ij} \left[\frac{2\kappa}{1-\exp(-2\kappa)}\right] a^3_s(Q_2^*) , \label{pmceq}
\end{eqnarray}
where $\kappa=\frac{\tilde{D}_{2ij}}{D_{1ij}} a_s(Q_2^*) + \frac{F_{2ij}}{D_{1ij}} \left(\frac{\pi}{v} \right) a_s(Q_2^*)$. Here $Q^{*}_{1}$ and $Q^{**}_{1}$ are the LO and NLO PMC scales for the non-Coulomb part, and $Q^{*}_2$ is the LO PMC scale for the Coulomb part. The PMC coefficients and PMC scales, together with their detailed derivations, can be found in Ref.~\cite{PMCStanWu}.

When we do numerical calculations, the input parameters are chosen with the following values: for the top quark mass, we adopt the PDG value~\cite{pdg}; i.e. $m_t = 172.9\pm1.1$ GeV. For the PDFs, we adopt the CTEQ CT10 set with $\alpha_s(m_Z)= 0.118$~\cite{cteq}. Our results for the $t \bar{t}$ production cross-sections are presented in Table \ref{scaleun} where the total cross-sections which are derived by using the PMC scale setting and the conventional scale setting are presented.

\begin{widetext}
\begin{center}
\begin{table}
\begin{tabular}{|c||c|c|c|c|c||c|c|c|}
\hline
& \multicolumn{5}{c||}{PMC scale setting} & \multicolumn{3}{c|}{Conventional scale setting} \\
\hline
~~~ ~~~ & ~$Q=m_t/4$~ & ~$Q=m_t$~ & ~$Q=10\,m_t$~ & ~$Q=20\,m_t$~& ~$Q=\sqrt{s}$~ & ~$\mu_R\equiv m_t/2$~ & ~$\mu_R\equiv m_t$~ & ~$\mu_R\equiv 2\,m_t$~ \\
\hline
Tevatron (1.96~TeV) & 7.620(5) & 7.626(3) & 7.625(5) & 7.624(6) & 7.628(5) & 7.742(5) & 7.489(3) & 7.199(5) \\
\hline
LHC (7~TeV) & 171.6(1) & 171.8(1) & 171.7(1) & 171.7(1) & 171.7(1) & 168.8(1) & 164.6(1) & 157.5(1) \\
\hline
LHC (14~TeV)& 941.8(8) & 941.3(5) & 942.0(8) & 941.4(8) & 942.2(8) & 923.8(7) & 907.4(4) & 870.9(6) \\
\hline
\end{tabular}
\caption{Dependence of the $t \bar{t}$ production cross-sections (in unit: pb) at the Tevatron and LHC on the initial renormalization scale $\mu_R^{\rm init}=Q$. Here $m_t=172.9$ GeV. The number in parenthesis shows the Monte Carlo uncertainty in the last digit. } \label{scaleun}
\end{table}
\end{center}
\end{widetext}

It is found that after PMC scale setting, the resulting total cross-sections for five disparate initial scales are equal to each other within part per mill accuracy \footnote{There is some small residual initial-scale dependence in the PMC scales because of unknown-higher-order $\{\beta_i\}$-terms.}. For comparison, we also present the results with conventional scale setting in Table \ref{scaleun}. For $\mu_R \in[m_t/2,2m_t]$, we obtain the usual renormalization scale-uncertainty $\left({}^{+3\%}_{-4\%}\right)$. This shows that the renormalization scale uncertainty is greatly suppressed and essentially eliminated using PMC even at the NNLO level. This is consistent with renormalization group invariance: there should be no dependence of the prediction for a physical observable on the choice of the initial renormalization scale.

\begin{figure}
\includegraphics[width=0.33\textwidth]{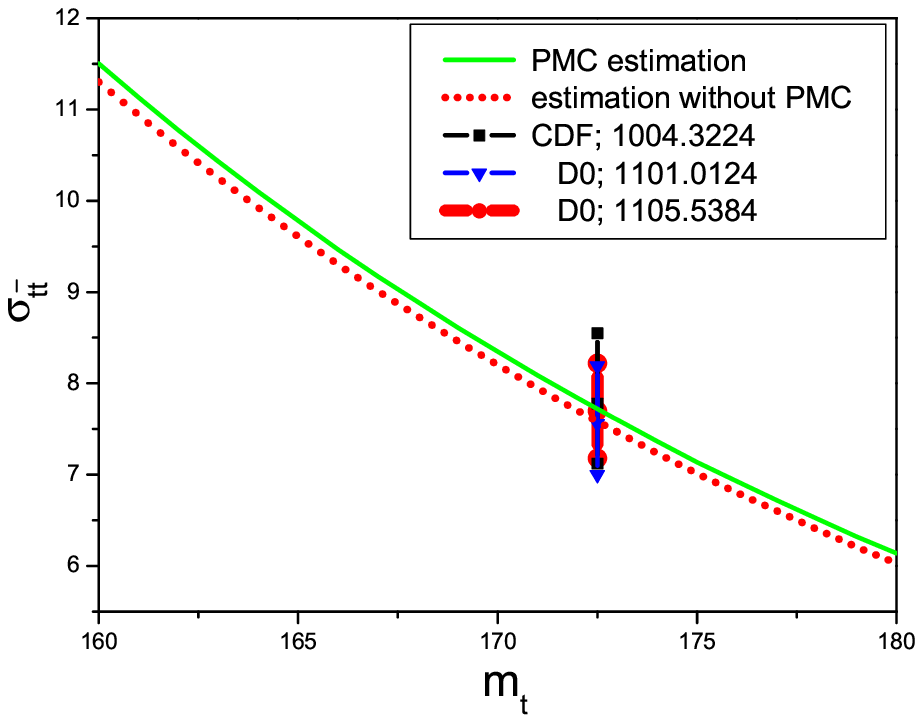}
\includegraphics[width=0.33\textwidth]{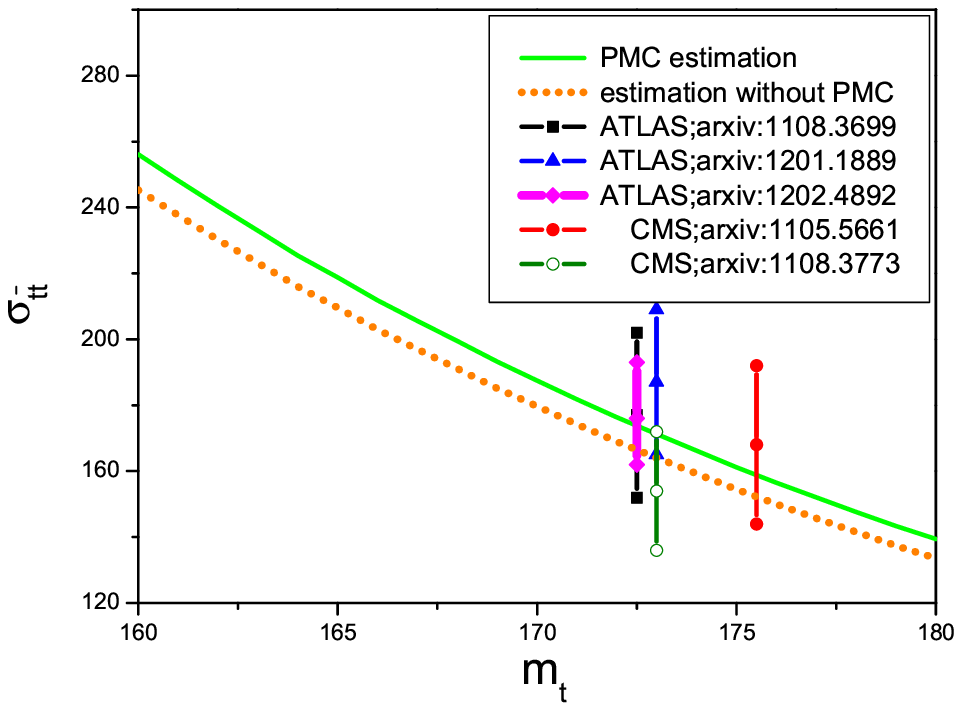}
\caption{Total cross-section $\sigma_{t\bar{t}}$ for the top quark pair production versus top quark mass. }
\label{mass}
\end{figure}

The PMC predictions for total cross-section $\sigma_{t\bar{t}}$ are sensitive to the top quark mass. We present $\sigma_{t\bar{t}}$ as a function of $m_t$ in Fig.(\ref{mass}). After PMC scale setting, the value of $\sigma_{t\bar{t}}$ becomes very close to the central values of the experimental data~\cite{cdf,d0,atlas,cms}. By varying $m_t=172.9\pm 1.1$ GeV~\cite{pdg}, we predict
\begin{eqnarray}
 \sigma_{\rm Tevatron,1.96\,TeV} &=& 7.626^{+0.265}_{-0.257} \;{\rm pb}\\
 \sigma_{\rm LHC,7\,TeV} &=& 171.8^{+5.8}_{-5.6} \;{\rm pb}\\
 \sigma_{\rm LHC,14\,TeV} &=& 941.3^{+28.4}_{-26.5}\;{\rm pb}
\end{eqnarray}

\begin{widetext}
\begin{center}
\begin{figure}[ht]
\includegraphics[width=0.32\textwidth]{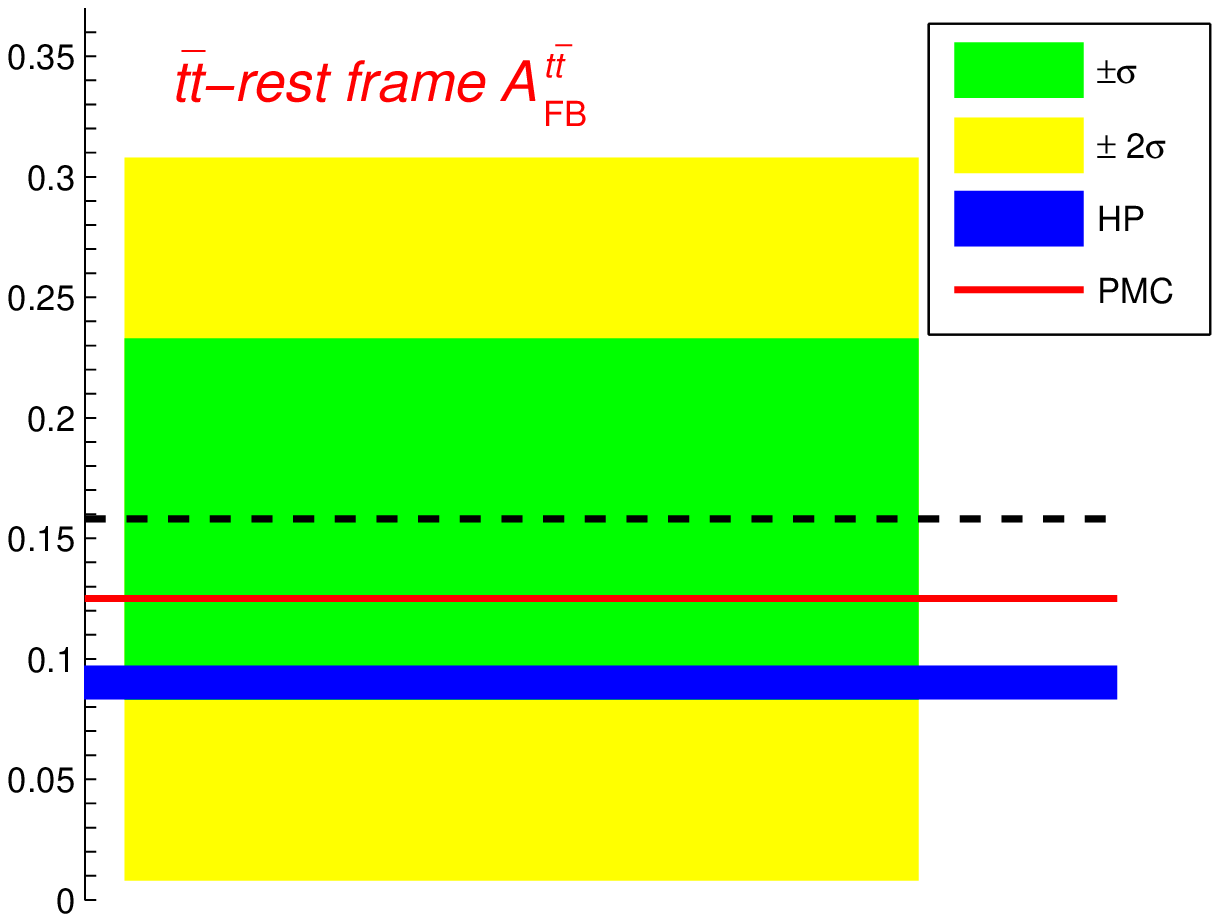}
\includegraphics[width=0.32\textwidth]{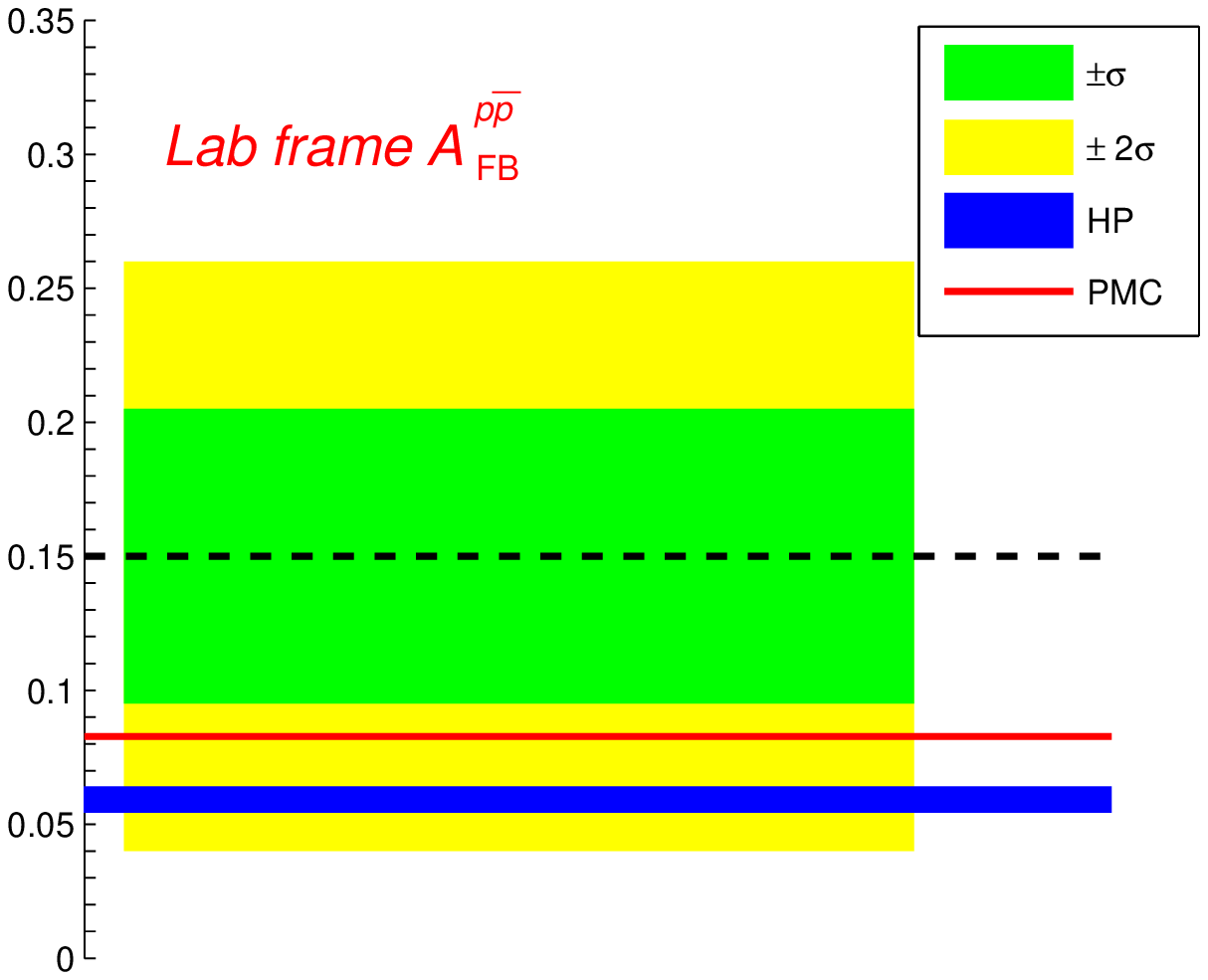}
\includegraphics[width=0.32\textwidth]{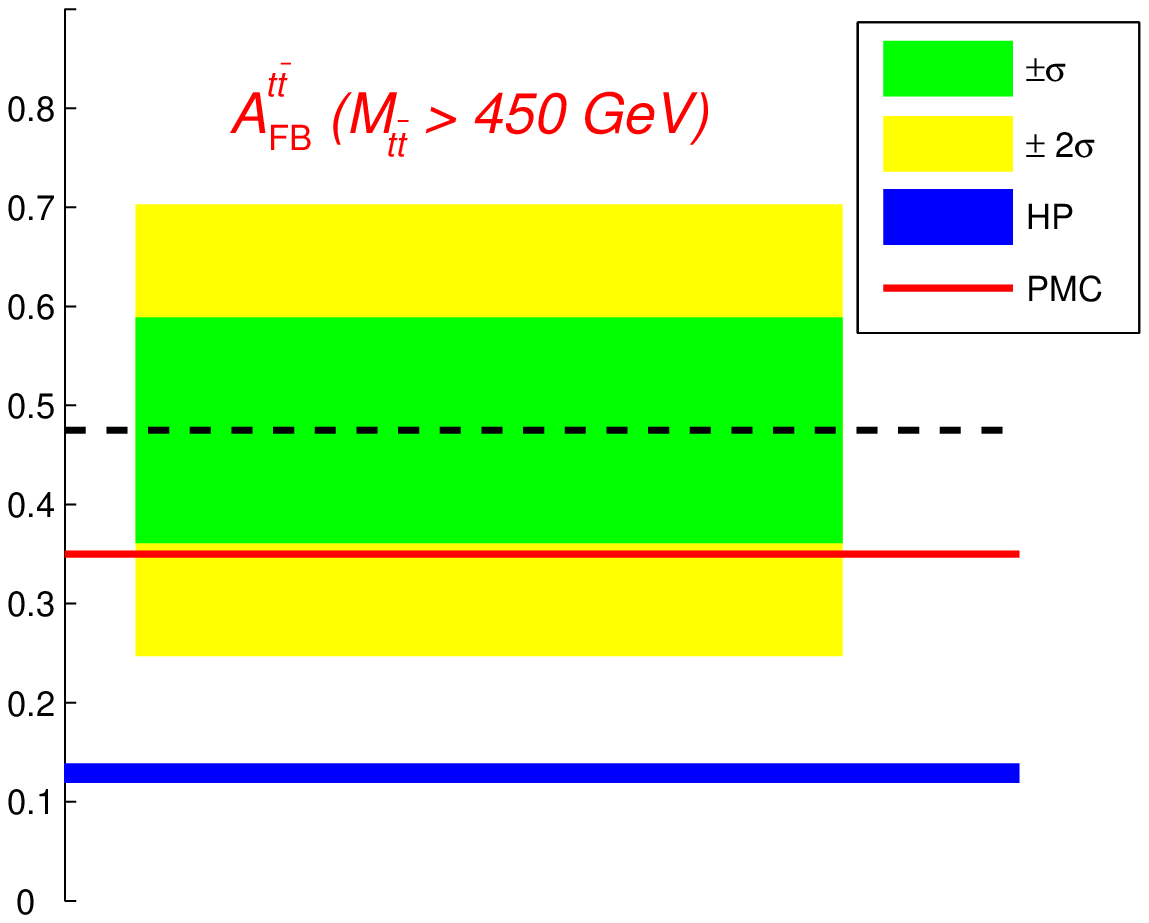}
\caption{Comparison of the PMC prediction with the CDF data~\cite{cdf2} for the $t\bar{t}$-pair forward-backward asymmetry for the whole phase-space. The left diagram is for $A_{FB}^{t\bar{t}}$ in the $t\bar{t}$-rest frame, the middle diagram is for $A_{FB}^{p\bar{p}}$ in the laboratory frame, and the right diagram is for $A_{FB}^{t\bar{t}}(M_{t\bar{t}}>450\; {\rm GeV})$. The Hollik and Pagani's results (HP)~\cite{qedc2} using conventional scale setting are presented for a comparison. The result for D0 data~\cite{d02} shows a similar behavior. } \label{pmcasy}
\end{figure}
\end{center}
\end{widetext}

We have recently shown that the large discrepancy between the Standard Model estimates using conventional scale setting and the CDF and D0 data~\cite{cdf2,d02} for the $t\bar{t}$-pair forward-backward asymmetry is mainly caused by improper setting of renormalization scale~\cite{topasymmetry}. After PMC scale setting, it is found that the NLO PMC scale has a dip behavior for the dominant asymmetric $(q\bar{q})$-channel; the importance of this channel to the asymmetry is thus increased. Then, after PMC scale setting, the $t\bar{t}$-pair forward-backward asymmetries $A_{FB}^{t\bar{t}}$ and $A_{FB}^{p\bar{p}}$ at the Tevatron are increased by $42\%$ in comparison with the previous estimates obtained by using conventional scale setting. We obtain $A_{FB}^{t\bar{t}} \simeq 12.5\%$, $A_{FB}^{p\bar{p}} \simeq 8.28\%$ and $A_{FB}^{t\bar{t}}(M_{t\bar{t}}>450 \;{\rm GeV}) \simeq 35.0\%$~\cite{topasymmetry}. These predictions have a $1\sigma$-deviation from the present CDF and D0 measurements; the large discrepancy of the top quark forward-backward asymmetry between the Standard Model estimate and the data are thus greatly reduced. This large improvement is explicitly shown in Fig.(\ref{pmcasy}), where Hollik and Pagani's results, which are derived under conventional scale setting~\cite{qedc2}, are presented for comparison.

{\bf Summary:} By using PMC scale setting, one obtains a unique, scale-fixed, scheme-independent prediction at any finite order in a systematic way. Since the renormalization scale and scheme ambiguities are removed, this procedure improves the precision of tests of the Standard Model and enhances the sensitivity to new phenomena. The PMC can be applied to a wide-variety of perturbatively-calculable collider and other processes.

We have applied PMC to study the $t\bar{t}$ hadroproduction cross-section $\sigma_{t\bar{t}}$ up to NNLO. The resulting LO- and NLO- terms are conformally invariant and scheme-independent, and the non-conformal contributions in the NNLO-terms are greatly suppressed. The PMC prediction for $\sigma_{t\bar{t}}$ agrees well with the present Tevatron and LHC data. We also verify that the initial renormalization scale-independence of the PMC prediction is satisfied to high accuracy at the NNLO: the total cross-section remains almost unchanged even when taking very disparate initial scales $\mu^{\rm init}_R$ equal to $m_t$, $10\,m_t$, $20\,m_t$, $\sqrt{s}$. The optimized PMC scales substantially eliminates the large discrepancy between the Standard Model estimation and the Tevatron data for the $t\bar{t}$-pair forward-backward asymmetry.

{\bf Acknowledgements:} We thank Leonardo di Giustino, Robert Shrock, Stefan Hoeche and Andrei Kataev for helpful conversations. This work was supported in part by the Program for New Century Excellent Talents in University under Grant NO.NCET-10-0882, Natural Science Foundation of China under Grant NO.11075225, and the Department of Energy contract DE-AC02-76SF00515. SLAC-PUB-14898.

\end{document}